\documentclass{article}
\usepackage{amsmath,epsfig}
\usepackage[preprint]{spconfa4}
\usepackage{xcolor}
\usepackage{cite}

\usepackage{graphicx}
\usepackage{stfloats}
\usepackage{times}
\usepackage{epsfig}
\usepackage{graphicx}
\usepackage{amsmath}
\usepackage{amssymb}
\usepackage{multirow}
\usepackage{amsthm,amsmath,amssymb}
\usepackage{mathrsfs}
\usepackage{booktabs}
\usepackage{hyperref}

\usepackage{lipsum}

\usepackage{setspace}
 
\newcommand\blfootnote[1]{%
  \begingroup
  \renewcommand\thefootnote{}\footnote{#1}%
  \addtocounter{footnote}{-1}%
  \endgroup
 }

\renewcommand{\thefootnote}{}

\let\OLDthebibliography\thebibliography
\renewcommand\thebibliography[1]{
  \OLDthebibliography{#1}
  \setlength{\parskip}{0pt}
  \setlength{\itemsep}{0pt plus 0.3ex}
}

\begin{document}\sloppy

\def\x{{\mathbf x}}
\def\L{{\cal L}}

\title{HarmoF0: Logarithmic Scale Dilated Convolution For Pitch Estimation}
%
\name{\large Weixing Wei$^{1}$, Peilin Li$^{1}$, Yi Yu$^{\ast2}$  and Wei Li$^{\ast1,3}$}
\address
{\normalsize$^{1}$ School of Computer Science and Technology, Fudan University, Shanghai, China
\\
\normalsize$^{2}$ Digital Content and Media Sciences Research Division, National Institute of Informatics (NII), Tokyo, Japan
\\\normalsize$^{3}$Shanghai Key Laboratory of Intelligent Information Processing, Fudan University, Shanghai, China
\\\normalsize wxwei20@fudan.edu.cn, plli21@m.fudan.edu.cn, yiyu@nii.ac.jp, weili-fudan@fudan.edu.cn
}

\maketitle

\begin{abstract}

\blfootnote{*Yi Yu and Wei Li are the corresponding authors.}
Sounds, especially music, contain various harmonic components scattered in the frequency dimension. It is difficult for normal convolutional neural networks to observe these overtones. This paper introduces a multiple rates dilated causal convolution (MRDC-Conv) method to capture the harmonic structure in logarithmic scale spectrograms efficiently. The harmonic is helpful for pitch estimation, which is important for many sound processing applications. We propose HarmoF0, a fully convolutional network, to evaluate the MRDC-Conv and other dilated convolutions in pitch estimation. The results show that this model outperforms the DeepF0,
yields state-of-the-art performance in three datasets, and simultaneously reduces more than 90\% parameters. We also find that it has stronger noise resistance and fewer octave errors. The code and pre-trained model are available at \url{https://github.com/WX-Wei/HarmoF0}.

\end{abstract}

\begin{keywords}
dilated convolution, pitch estimation, harmonic model, sound signals processing
\end{keywords}

\section{Introduction}
\label{sec:intro}




As a basic property of mono source sound, fundamental frequency (f0) is important for audio signal analysis. Pitch is an auditory sensation closely related to f0. They are not exactly equivalent but are often seen as the same concept in audio signal processing. Fundamental frequency estimation or pitch estimation is a very important technology for many audio-based applications such as speech analysis and music information retrieval \cite{MIR_review_2018}.

It is common to feed spectrograms to convolutional neural networks (CNN).
However, processing spectrograms as images in CNN models has some problems. Spectrograms lose the phase information. The energy of audio events spans the frequency dimension and overlaps with each other, making it difficult for CNN to extract features \cite{wyse2017audio_spec}. Some researchers try to use the raw waveform as the input of the neural networks. The DeepF0 \cite{DeepF0_singh2021} feeds the raw audio sample to the model and applies dilated convolution \cite{DBLP_2016YuK15} to estimate pitch. It achieved high performance, but the CNN model working in the time domain needs large kernel sizes to keep local information as well as expand the receptive field. That makes the model large and computationally intensive.


Real sounds usually contain a series of overtones, integral multiples of the fundamental frequency.
Humans perceive the pitch of complex tones not just according to the energy of fundamental frequency but the entire harmonic series \cite{how_we_hear_oxenham2018}. In this work, we attempt to design a model to capture the harmonic series and simulate such a mechanism.

In the spectrograms, the harmonic series is sparse in the frequency domain
, which needs to stack multiple convolution layers to cover the harmonic series. Dilated convolution \cite{DBLP_2016YuK15} is a way to expand the receptive field rapidly.
By setting a specific dilation rate, the receptive field can be increased at an exponential rate.
But with a fixed dilation rate, dilated convolution can not leverage the integer multiple relationships between fundamental frequency and overtones in the linear scale.
In view of this problem, we propose a dilated convolution method with multiple dilation rates in a logarithmic scale. And based on this approach, we construct HarmoF0, a fully convolutional network for pitch estimation.
Empirical evaluation demonstrates that our model outperforms the state-of-the-art DeepF0 \cite{DeepF0_singh2021}, but with much fewer parameters. In addition, in noisy environments, it still surpasses all the baselines, including DeepF0, SWIPE \cite{SWIPE_camacho2008}, pYin \cite{pYIN_mauch2014}, and CREPE \cite{CREPE_kim2018}, revealing a high capability of noise resistance.

The rest of this paper is organized as follows. Section 2 introduces the related works. Section 3 describes the proposed multiple rates dilated convolution and HarmoF0 model for pitch estimation. Section 4 illustrates the experimental setup, with results analyzed in Section 5. Conclusions are discussed in Section 6.

\section{Related works}
\label{sec:related}




In the past decades, plenty of algorithms are devised for monotonic pitch estimation. Traditional methods are based on digital signal processing (DSP), working in the time domain (RAPT \cite{RAPT_talkin1995}, YIN \cite{YIN_de2002}, and pYIN \cite{pYIN_mauch2014}), in the frequency domain (SWIPE \cite{SWIPE_camacho2008}
) or combining the two \cite{kawahara2005nearly}, usually applying a certain candidate-generating function to select target pitch. These algorithms can achieve quite accurate results for simple mono source sounds. However, they have false estimation problems in noisy conditions, especially the ``octave error'' \cite{melody_challenges_salamon2014}. With traditional signal processing methods, it is difficult to reduce ``multiple-frequency errors'' and ``half-frequency errors" at the same time: efforts to reduce one often result in an increase in the other. Post-processing employing curve smoothing methods  \cite{pYIN_mauch2014}\cite{melody_contour_salamon2012}  to track the best possible candidate pitch can reduce such errors but are computationally intensive and not robust to noise.

In recent years, with the advance of deep learning and highly annotated data, data-driven approaches for pitch estimation surpass the hand-crafted signal processing methods. The CREPE \cite{CREPE_kim2018} applied a convolutional neural network on the waveforms to estimate the pitch, and outperformed the pYIN on two re-synthesizing datasets. But the CREPE model, with parameters up to 22.2M, is computationally expensive. To reduce the parameters and speed up the calculation, the DeepF0 \cite{DeepF0_singh2021} is proposed by applying dilated convolution and skip connections, achieving the state-of-the-art on the tested datasets. However, as a compromise, the DeepF0 does not perform as well as the CREPE in loud noisy environments.
There are also efforts to estimate pitch using self-supervised learning. The SPICE \cite{SPICE_gfeller2020} algorithm, while using relative pitch for self-supervised learning, has a complicated training process.

Deep learning based methods have made significant progress in pitch estimation, but it still needs to be further explored in reducing model sizes and improving robustness in complex acoustic scenes.

\section{Architecture}
\label{sec:arch}

In this section, we first introduce the proposed multiple rates causal dilated convolution. Then we present the detailed structure of the network.



%

\subsection{Multiple rates dilated convolution}

\subsubsection{Harmonics in a logarithmic scale}

\begin{figure}[h]
\centering
\includegraphics[scale=0.47]{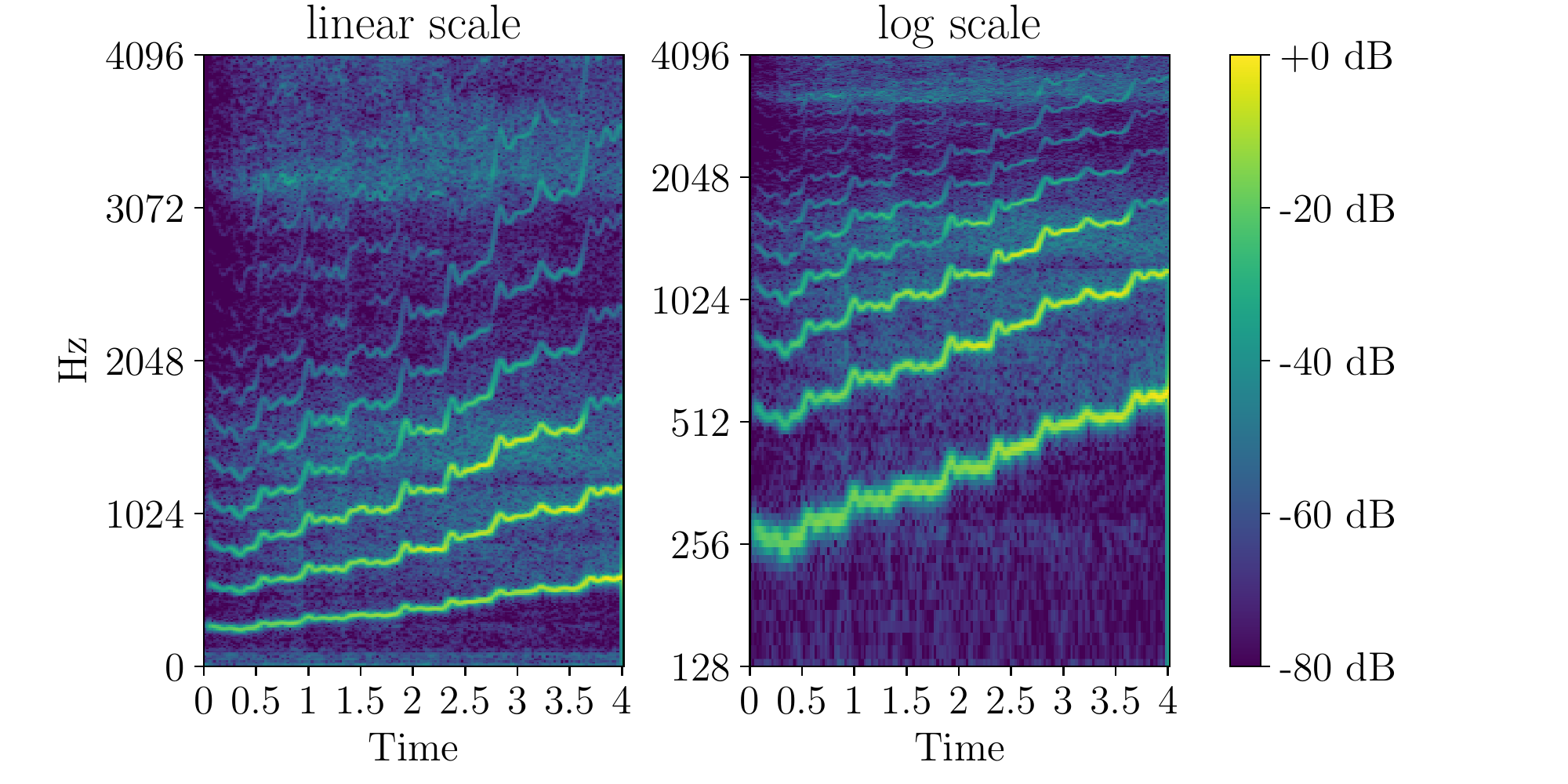}
\caption{The linear scale and log scale spectrograms of a singing voice. The intervals between the harmonic waves are different in the linear scale spectrum but are congruous in the log scale spectrum. }
\label{fig:scale_spectrum}
\end{figure}

Fig.\ref{fig:scale_spectrum} illustrates the intervals of harmonic series in linear-scale and log-scale spectrograms. The interval between adjacent harmonics changes as the f0 changes in a linear-frequency scale. This is not conducive to determining the dilation rate. In the logarithmic scale, for a harmonic series with arbitrary f0, the interval between adjacent harmonics becomes a constant value. As shown in Eq.(\ref{eq:log_intervals}), with the base of $2^{\frac{1}{Q}}$ where $Q=48$ (number of bins per octave), the number of harmonic series $N_{har}=5$, the intervals of adjacent harmonics in a harmonic series are rounded to integers as follows: 48, 28, 20, 15.
It is intuitive to set the dilation rate as these values to capture harmonic series.

\vspace{-0.5cm}

\begin{equation}
\begin{aligned}
d_{k} &= log_{2^{1/Q}} (f_0 \cdot (k+1)) - log_{2^{1/Q}} (f_0 \cdot k) \\
&= Q \cdot log_{2}(\frac{k+1}{k})
\end{aligned}
\label{eq:log_intervals}
\end{equation}
Where $k$ is the serial number of the harmonic series, $d_{k}$ denotes the interval of adjacent harmonics, $Q$ indicates the number of bins per octave, and $f_0$ is fundamental frequency.


\subsubsection{Multiple rates dilated causal convolution}

\label{sec:MRDC-Conv}

\begin{figure}[b]
\centering
\includegraphics[scale=0.25]{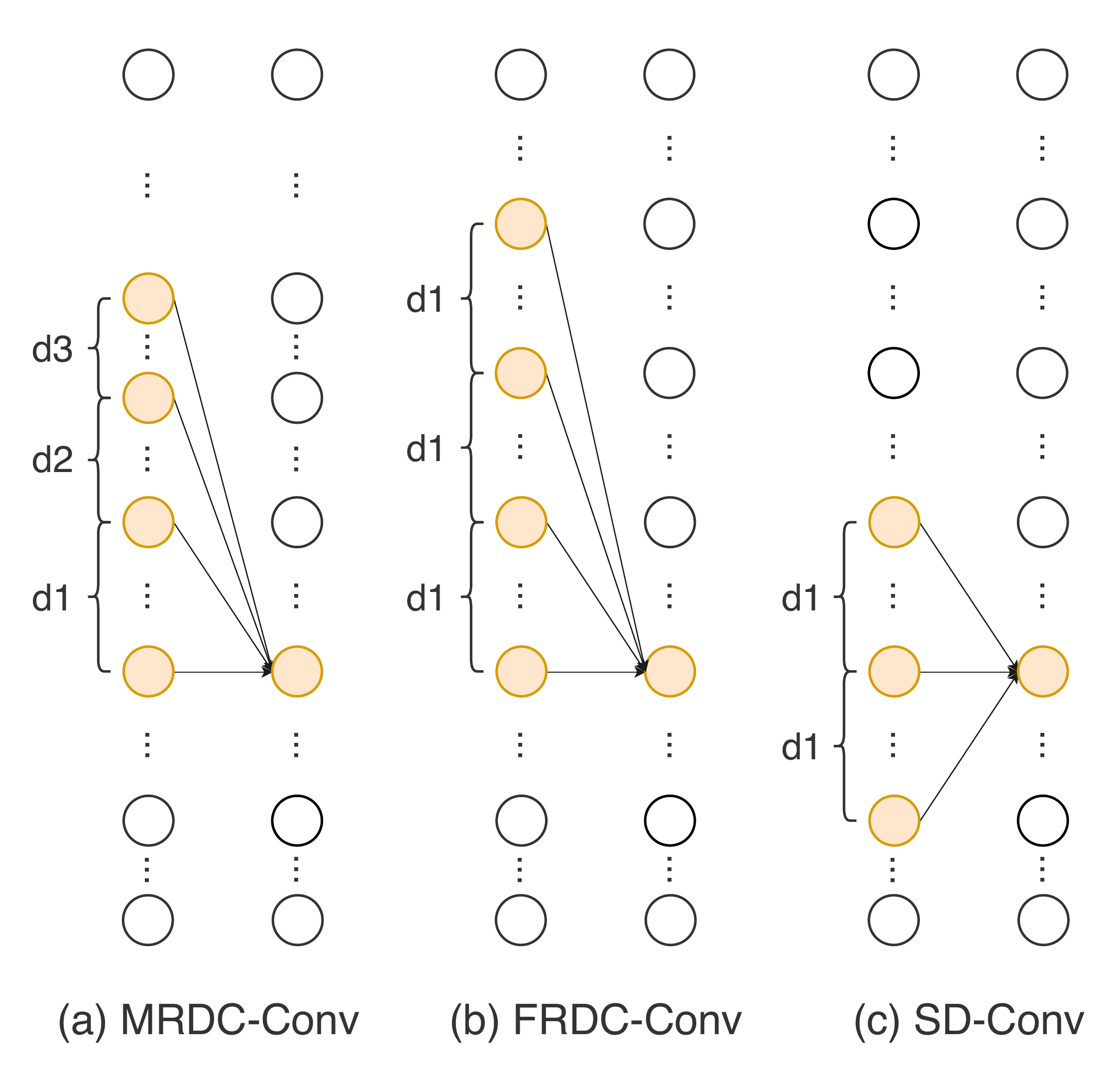}
\caption{Three types dilated convolution used in this work. }
\label{fig:dilated-conv-types}
\end{figure}

The normal dilated convolution has a fixed dilation rate for concurrent computing. To apply multiple dilation rates in the same dilated convolution, we propose the Multiple Rates Dilated Causal Convolution (MRDC-Conv), as shown in the Figure \ref{fig:dilated-conv-types} (a). The dilation rates $d_1, d_2, and\ d_3$ are calculated by the Eq.(\ref{eq:log_intervals}).
The MRDC-Conv is causal \cite{wavenet}, ensuring that the current output is derived from one side of the sequences.
Figure  \ref{fig:dilated-conv-types} (b) and (c) are the Fixed Rate Dilated Causal Convolution (FRDC-Conv) and Standard Dilated Convolution (SD-Conv) with fixed dilation rate $d_1$.



\begin{figure}[htb]
\centering
\includegraphics[scale=0.27]{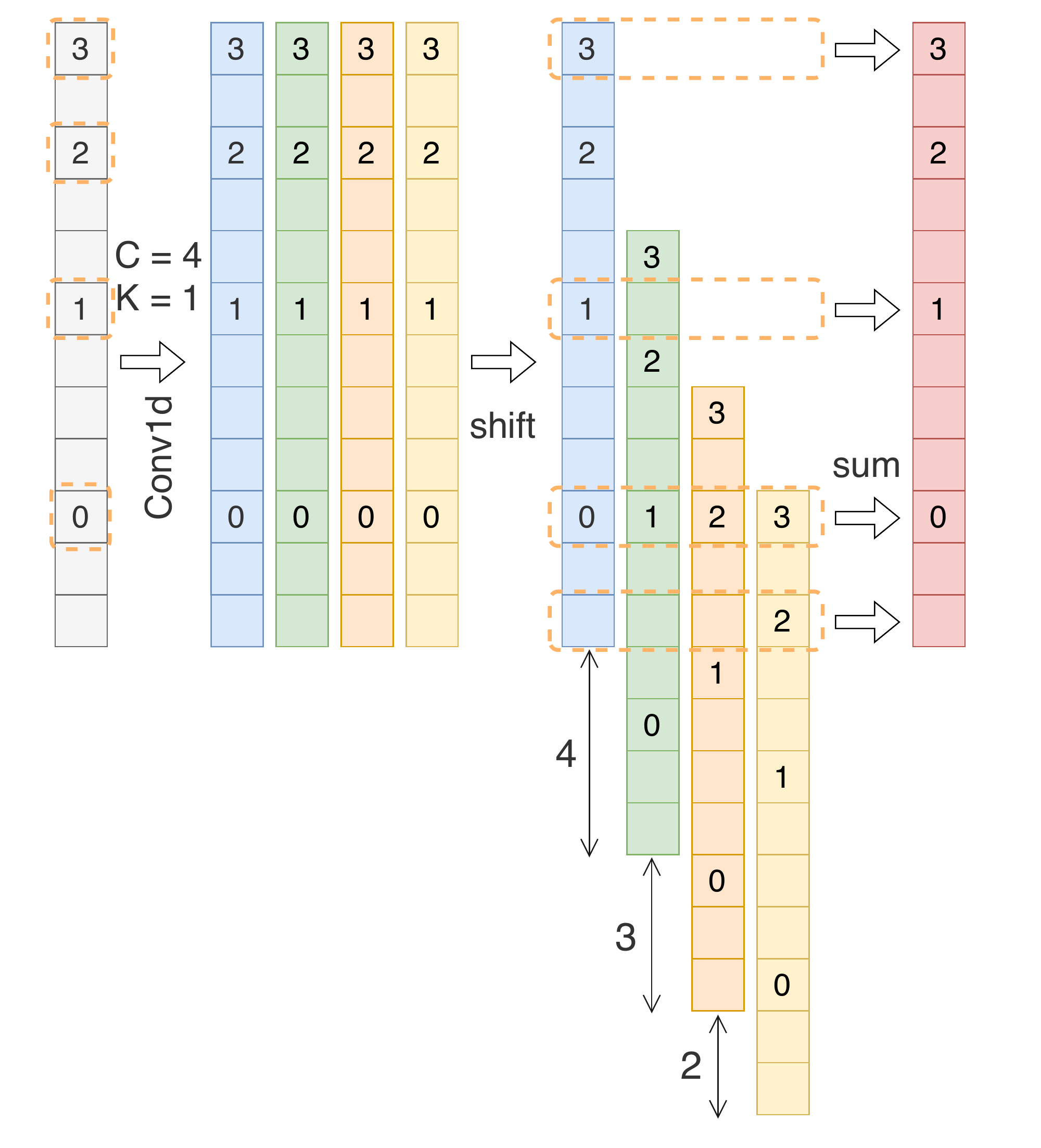}
\caption{Implementation of dilated causal convolution with multiple rates: 4, 3, and 2. }
\label{fig:dilated_conv}
\end{figure}

Fig. \ref{fig:dilated_conv} demonstrates the implementation of a one-dimension MRDC-Conv with dilation rates 4, 3, and 2. Firstly, the one-dimensional input sequence is convoluted with C=4 filters, each with kernel size K=1, to obtain four different sequences. Next, the four sequences are shifted vertically according to the dilation rates to the corresponding positions, and then the elements in the same horizontal column are summed. Elements outside the valid range are discarded. In this way, the multiple rate dilated convolution can be made in parallel.



\subsection{Architecture of HarmoF0}


Raw audios are resampled to 16 kHz and clipped to a fixed length of 2 seconds. Then the short-time Fourier transform is applied with a Hamming window of size 1024 and a hop size of 320 to compute spectrograms.
The frequency dimension is converted to a logarithmic scale with 352 frequency bins, which separate the frequency range from 27.5 Hz to 4371.3 Hz spaced 25 cents.
The logarithmic spectrograms with the size of (352, $T$) are used as the inputs for training, with frame number $T=100$.



\begin{figure}[htb]
\centering
\includegraphics[scale=0.45]{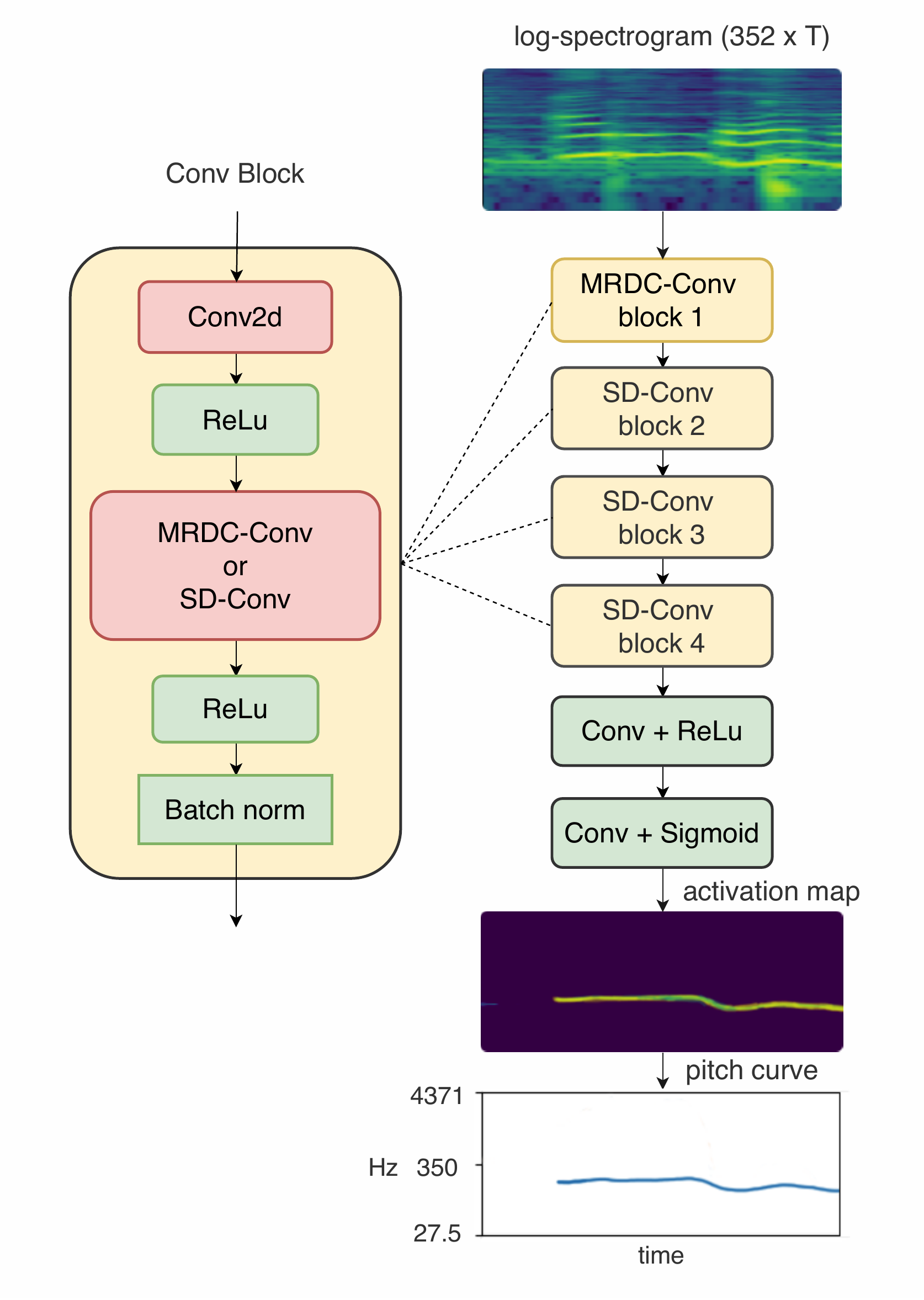}
\caption{The architecture of the HarmoF0.}
\label{fig:harmonic_net}
\end{figure}

The architecture of the proposed HarmoF0 model is shown in Fig. \ref{fig:harmonic_net}. It is a fully convolutional network, with four convolution blocks, two 1$\times$1 convolution layers, a ReLu, and a sigmoid activation function. The hyperparameters of all convolution layers are shown in Table \ref{tab:conv_parameters}. Each convolution block contains a normal convolution layer and a dilated convolution layer. Both convolution layers are followed by a ReLu activation. A batch normalization is applied at the last of each block. All dilated convolution operations in our model are performed only in the frequency domain dimension. The dilated convolution layer of the first block is an MRDC-Conv layer,  which has multiple dilation rates as described in Section \ref{sec:MRDC-Conv}. The dilation rates are calculated by the Eq. (\ref{eq:log_intervals}) with $Q=48$.  We set the number of harmonic series to $N_{har}=12$. The dilated convolution layers of the rest blocks are SD-Conv layers,  having a dilation rate of $d=48$. Following the last convolution block, two convolution layers with kernel size of 1$\times$1 are applied to decrease the channels from 128 to 1. Finally, an output is obtained with the same size of the input. The frequency bin with maximum activation value is considered as the estimated pitch of a frame. And results can be converted to frequency by Eq. (\ref{index_to_hz}).

\begin{equation}
f(i) = f_{min}  \cdot 2^{\frac{i}{Q}}  [Hz]
\label{index_to_hz}
\end{equation}
Here minimum frequency $f_{min}=27.5   Hz$ , and i is the frequency bin index, ranging from 0 to 351.\\

During training, we convert the ground-truth pitch of each frame to a 352 dimension one-hot vector $y$. Since the positive and negative categories are unbalanced, we use weighted cross entropy loss and set positive weight $w=20$ in the experiments. The model is trained to minimize the weighted binary cross-entropy loss between the target vector $y$ and the estimation one-hot vector $\hat{y}$ of each frame:

\vspace{-0.2cm}

\begin{equation}
\mathcal{L}(\mathbf{y}, \hat{\mathbf{y}})=\sum_{i=1}^{352}\left(-wy_{i} \log \hat{y}_{i}-\left(1-y_{i}\right) \log \left(1-\hat{y}_{i}\right)\right)
\end{equation}




\label{sec:Model parameters}

\vspace{0.2cm}

\begin{table}[htb]
\Huge
\renewcommand{\arraystretch}{1.2}
\centering

\resizebox{.85\columnwidth}{!}{
\begin{tabular}{llccr}
\toprule[3.5pt]
&       & filters & kernel size & \multicolumn{1}{c}{dilation rate} \\
\midrule[2.5pt]
\multicolumn{1}{c}{\multirow{2}[2]{*}{block 1}} & Conv2d & 32    & 3x3   & \multicolumn{1}{c}{-}    \\
& MRDC-Conv & 32    & 11x1  & \multicolumn{1}{c}{48, 28, 20, ... , 6} \\
\midrule
\multicolumn{1}{c}{\multirow{2}[2]{*}{block 2}} & Conv2d & 64    & 3x3   & \multicolumn{1}{c}{-}  \\
& SD-Conv & 64    & 3x1   & \multicolumn{1}{c}{48} \\
\midrule
\multicolumn{1}{c}{\multirow{2}[2]{*}{block 3}} & Conv2d & 128   & 3x3   & \multicolumn{1}{c}{-} \\
& SD-Conv & 128   & 3x1   & \multicolumn{1}{c}{48} \\
\midrule
\multicolumn{1}{c}{\multirow{2}[2]{*}{block 4}} & Conv2d & 128   & 3x3   & \multicolumn{1}{c}{-} \\
& SD-Conv & 128   & 3x1   & \multicolumn{1}{c}{48} \\
\midrule
Conv 5 & Conv2d & 64    & 1x1   & \multicolumn{1}{c}{-} \\
\midrule
Conv 6 & Conv2d & 1     & 1x1   & \multicolumn{1}{c}{-} \\
\bottomrule[3.5pt]
\end{tabular}%
}
\caption{Hyperparameters of convolution layers. The MRDC-Conv layer has 11 different dilation rates, while all SD-Conv layers have the same dilation rate of 48.}
\label{tab:conv_parameters}%
\end{table}%

\section{Experiments}
\label{sec:experiments}
In this section we introduce the datasets we used and the details of the experiment.
\subsection{Datasets}
Three different datasets are selected to train and evaluate the proposed model in our experiments.
\begin{itemize}
\item[$\bullet$] MIR-1K \cite{MIR-1K} is a dataset containing 1000 song clips with the music accompaniment and the singing voice recorded as left and right channels, respectively; and manual annotations of pitch. The total length of the dataset is 133 minutes.
\item[$\bullet$] MDB-stem-synth \cite{MDB-stem-synth} contains 230 solo stems (tracks) spanning a variety of musical instruments and
voices from the MedleyDB \cite{Bittner2014MedleyDBAM} dataset, and perfect f0 annotations of the stem (track) obtained via the analysis/synthesis method.
\item[$\bullet$] PTDB-TUG \cite{PTDB-TUG} is a speech database for pitch tracking. It contains 20 English native speakers and the extracted pitch trajectories as a reference. This database consists of 4720 recorded sentences.
\end{itemize}


\subsection{Experimental setup}

The experiments are carried out with 5-fold cross-validation, using a 60\%/20\%/20\% training, validation, and test split. The model is trained using Adam with default hyperparameters and a learning rate of $10^{-3}$ for 50 epochs. The batch size is set to 24.


\subsection{Evaluation Measures}

We use raw pitch accuracy (RPA) and raw chroma accuracy (RCA) \cite{melody_challenges_salamon2014} to measure the pitch estimation performance within a threshold of 50 cents.
The evaluation metrics are calculated using the mir\_eval\ \cite{Raffel14mir_eval:a} library.


\section{Results and discussion}
\label{sec:result}

The proposed model is compared with 4 state-of-the-art methods: pYin \cite{pYIN_mauch2014}, SWIPE \cite{SWIPE_camacho2008}, CREPE \cite{CREPE_kim2018}, and DeepF0 \cite{DeepF0_singh2021}. The performance in noisy conditions and ablation study are also discussed in this section.


\subsection{Accuracy compared with baselines}

The pitch estimation performance of the proposed model is depicted in Table \ref{tab:accuracy}. As can be seen in Table \ref{tab:accuracy}, our model achieved state-of-the-art accuracy in the three datasets. In MIR-1k and MDB-stem-synth, all RPA and RCA exceed 98\%. A significant advantage is that our model has much fewer parameters, which is around 0.377 million.
HarmoF0 is proposed to recognize harmonic distribution on spectrograms and estimates pitch based on harmonic series. Speech contains fewer harmonic components than musical sound. So it is easy to interpret that, in the speech dataset PTDB-TUG,  the model does not perform as well as in the previous two datasets. But with RPA of 93.56\% and RCA of 94.59\%, it still outperforms the other four models.

\begin{table}[h]
\LARGE
\begin{center}
\resizebox{.99\columnwidth}{!}{
\setlength{\arrayrulewidth}{0.3mm}
\renewcommand{\arraystretch}{1.1}
\begin{tabular}{c|c|c|c|c|c}

\toprule[3pt]
\multirow{2}{*}{Model} & \multirow{2}{*}{Params} & \multirow{2}{*}{Metrics} & \multicolumn{3}{c}{Datasets}\\

\cline{4-6}

&&&  \LARGE \quad MIR-1k \quad  &\LARGE \ MDB-stem & \LARGE PTDB-TUG\\
\midrule[2.5pt]
\multirow{2}{*}{pYIN \cite{pYIN_mauch2014}} & \multirow{2}{*}{-} & RPA (\%) & 90.47 & 90.12 & 50.51 \\

&& RCA (\%) & 91.06 & 90.71 & 51.30 \\
\hline

\multirow{2}{*}{SWIPE \cite{SWIPE_camacho2008}} & \multirow{2}{*}{-} & RPA (\%) & 96.36 & 92.50 & 67.45 \\

&& RCA (\%) & 96.73 & 93.34 & 69.50 \\

\hline
\multirow{2}{*}{CREPE \cite{CREPE_kim2018}} & \multirow{2}{*}{22.2M} & RPA (\%) & 96.41 & 96.34 & 81.44 \\
&& RCA (\%) & 96.72 & 96.74 & 84.26 \\
\hline
\multirow{2}{*}{DeepF0 \cite{DeepF0_singh2021}} & \multirow{2}{*}{5M} & RPA (\%) & 97.82  & 98.38  & 93.14 \\
&& RCA (\%) & 98.28  & 98.44 & 93.47 \\
\hline
\multirow{2}{*}{HarmoF0} & \multirow{2}{*}{0.377M } & RPA (\%) & \pmb{98.34} & \pmb{98.40} & \pmb{93.56} \\
&& RCA (\%) & \pmb{98.46} & \pmb{98.46} & \pmb{94.59} \\
\bottomrule[3pt]

\end{tabular}
}
\caption{Average raw pitch accuracy and raw chroma accuracy tested on three different test datasets.} \label{tab:accuracy}
\end{center}
\end{table}

\subsection{Performance in noisy conditions}
The performance in noisy environments is related to the robustness and the application in real situations. For this reason, we examine the noise robustness of the model. We mix musical accompaniments to the clean sounds in the MIR-1k dataset with different signal-to-noise ratio (SNR) values: 0, 10, and 20 dB. The results on the MIR-1k dataset are shown in Table \ref{tab:noise}. As can be seen, HarmoF0 shows better performance in noisy situations, especially in 0dB. In addition, in the RPA aspect, HarmoF0 has the best performance among the five models, achieving RPA of 85.11\% even when SNR = 0dB. The RPA and RCA are closer in HarmoF0 results, indicating fewer octave errors.

\vspace{0.8cm}
\begin{table}[htbp]
\footnotesize
\begin{center}
\resizebox{0.98\columnwidth}{!}{
\setlength{\arrayrulewidth}{0.3mm}
\renewcommand{\arraystretch}{1.1}

\begin{tabular}{c|c|c|c|c|c}

\toprule[1.5pt]

\multirow{2}{*}{Model} & \multirow{2}{*}{Metrics} & \multicolumn{4}{c}{Noise Profile}\\
\cline{3-6}
&&Clean & 20dB & 10dB & 0dB\\
\midrule[1.2pt]
\multirow{2}{*}{pYIN \cite{pYIN_mauch2014}} & RPA (\%) & 90.47 & 89.82 & 74.10 & 14.70 \\
& RCA (\%) & 91.06 & 90.38 & 80.60 & 33.78 \\
\hline
\multirow{2}{*}{SWIPE \cite{SWIPE_camacho2008}} & RPA (\%) & 96.36  & 94.44 & 83.55 & 39.44 \\
& RCA (\%) & 96.73 & 94.98 & 86.12 & 48.33 \\
\hline
\multirow{2}{*}{CREPE \cite{CREPE_kim2018}} & RPA (\%) & 96.41  & 96.17 & 94.51 & 84.22 \\
& RCA (\%) & 96.72 & 96.61 & 95.57 & 87.16 \\
\hline
\multirow{2}{*}{DeepF0 \cite{DeepF0_singh2021}} & RPA (\%) & 97.82 & 97.39 & 94.77 & 79.52 \\
& RCA (\%) & 98.28 & \pmb{98.09}  & \pmb{96.35} & 84.37 \\
\hline
\multirow{2}{*}{HarmoF0} & RPA (\%) & \pmb{98.34} & \pmb{97.82} & \pmb{95.56} & \pmb{85.11} \\
& RCA (\%) & \pmb{98.46} & 97.98 & 96.14 & \pmb{88.11} \\
\bottomrule[1.5pt]
\end{tabular}
}
\caption{Average raw pitch accuracy and raw chroma accuracy on the MIR-1k dataset with various levels of SNR.} \label{tab:noise}
\end{center}
\end{table}

\subsection{Ablation study}
\begin{figure}[htbp]
\includegraphics[width=8cm]{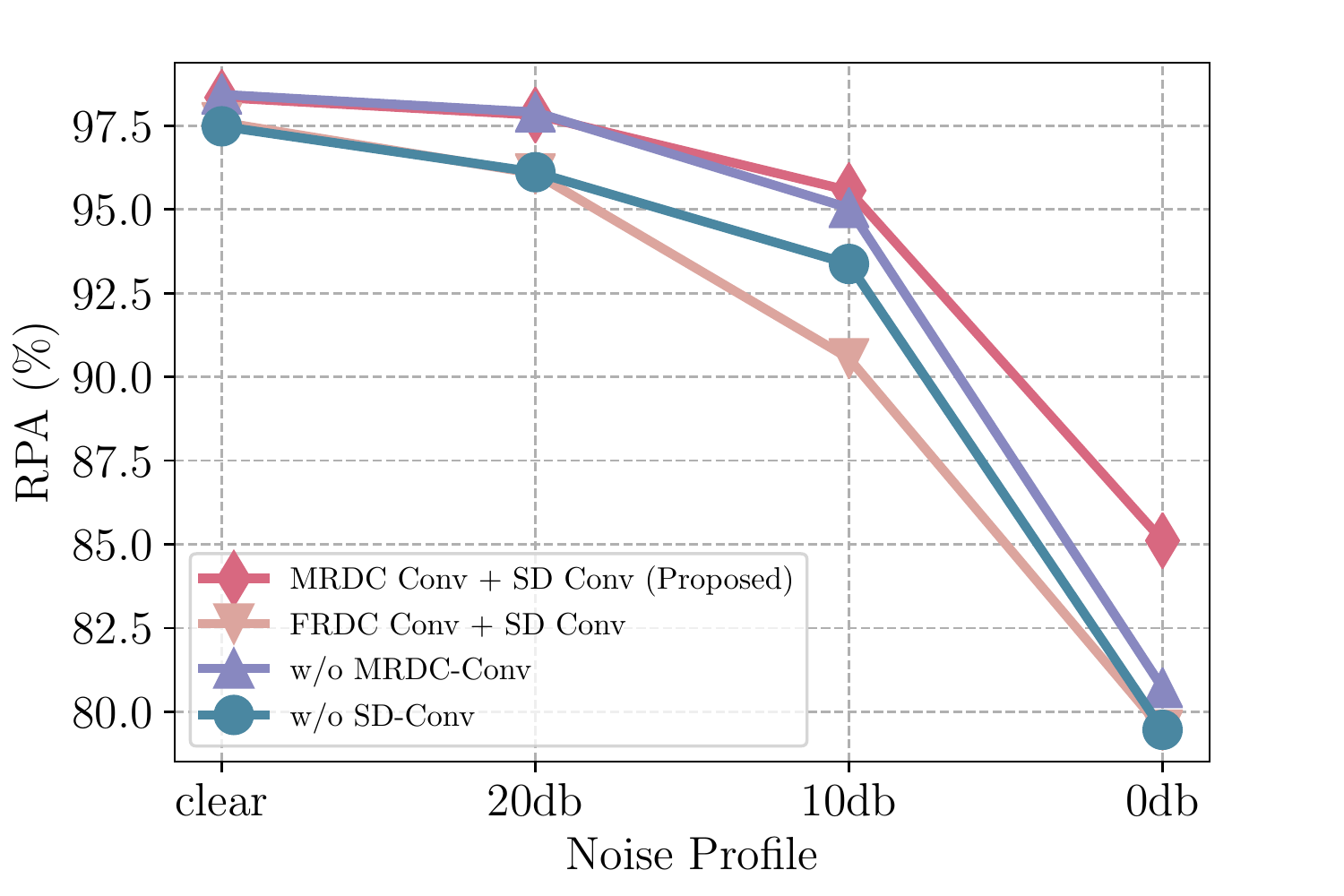}
\caption{The RPA performances of models with different dilated convolution under different noise levels in the MIR-1k dataset. }
\label{fig:ablation}
\end{figure}

Here, we further analyze the role of the MRDC-Conv layer and SD-Conv layer in HarmoF0. In addition to the proposed model, we evaluate the other three variants of our proposed model: replacing the MRDC-Conv layer in block 1 by a FRDC-Conv layer with dilation rate $d=48$, replacing the MRDC-Conv layer with a standard CNN layer, and replacing the SD-Conv layers with standard CNN layers, while keeping other hyperparameters the same. And the results are presented in Fig. \ref{fig:ablation}.

The proposed model outperforms all the other three ablated models. As the SNR becomes smaller, the advantage of the proposed model becomes more apparent. Particularly, when SNR is 0db, the proposed model shows RPA with 85.11\%,  while the maximum of the other three models is 80.96\%, with a gap of 4.15\%. It indicates that both the MRDC-Conv layer and SD-Conv layer are critical for the robustness of the model.
Humans perceive the pitch of complex tones according to the energy of the harmonic series. The proposed model works in a similar way, as the MRDC-Conv enables the model to track a long harmonic series and estimate pitch by considering all the harmonics. Noisy sounds do not have similar patterns as target harmonics. Consequently, the model performs better in a noisy environment.


\section{Conclusion}
\label{sec:conclusion}

In this paper, we implemented a multiple rate causal dilation convolution (MRDC-Conv) to set customized dilation rates in the log-frequency dimension.
Based on the MRDC-Conv, we proposed HarmoF0 for pitch estimation. Our experimental results demonstrate that the proposed model outperforms the existing state-of-the-art algorithms on three datasets. The number of parameters is also reduced to less than 10\% compared to the DeepF0. The model sets dilation rates equal to the intervals of adjacent harmonics, making it extract information from spectrograms efficiently with smaller kernel sizes. It follows parameters reduced. HarmoF0 also shows low octave errors and better performance in noisy situations, indicating its promising applications in practical situations.

Although HarmoF0 is only tested in monophonic pitch estimation, it also has potential in melody extraction, polyphonic pitches tracking, and other music information retrieval fields. We will continue to improve the structure of the model, explore its application in other tasks, and evaluate its performance in future works.


\section{Acknoledgement}
This work was supported by National Key R\&D Program of China(2019YFC1711800), NSFC(62171138).



\begin{spacing}{0.94}

\bibliographystyle{ieeetr}
\bibliography{icme2022template}

\end{spacing}

\end{document}